\documentclass[amstex,twocolumn,showpacs,floats,floatfix,superscriptaddress,aps,pra]{revtex4}
\usepackage{amssymb}
\usepackage{amsmath}
\usepackage{calc}
\usepackage{graphicx}
\usepackage{bm}

\def\be{ \begin{equation} }
\def\ee{ \end{equation} }
\def\bea{ \begin{eqnarray} }
\def\eea{ \end{eqnarray} }
\def\bse{ \begin{subequations} }
\def\ese{ \end{subequations} }

\input{tcilatex}

\begin{document}

\begin{abstract}
This paper we propose how to apply the Dykhne-Davis-Pechukas (DDP) method
for optimization of adiabatic passage in a two-state system in the second
adiabtic basis.
\end{abstract}

\author{G. S. Vasilev}
\affiliation{Department of Physics, Sofia University, James Bourchier 5 blvd, 1164 Sofia,
Bulgaria}
\author{N. V. Vitanov}
\affiliation{Department of Physics, Sofia University, James Bourchier 5 blvd, 1164 Sofia,
Bulgaria}
\title{Superadiabatic optimization via Dykhne-Davis-Pechukas (DDP) method}
\date{\today }
\maketitle


\section{Introduction}


In this paper we propose how to utilize a recent idea of Guerin \emph{et al.}
\cite{Guerin} who applied the well-known Dykhne-Davis-Pechukas (DDP) method
\cite{Davis76} for optimization of adiabatic passage in a two-state system.
In order to adapt this approach to STIRAP, we reduce the three-level Raman
system to effective two-state systems in two limits: on exact resonance and
for large single-photon detuning. The optimization, which minimizes the
nonadiabatic transitions and maximizes the fidelity, leads to a particular
relation between the pulse shapes of the driving pump and Stokes fields.


\section{Optimization of adiabatic passage between two states \label%
{Sec-background}}

The probability amplitudes in a two-state system $\mathbf{c}(t)=\left[
c_{1}(t),c_{2}(t)\right] ^{T}$ satisfy the Schr\"{o}dinger equation,
\begin{equation}
\text{i}\hbar \frac{\text{d}}{\text{d}t}\mathbf{c}(t)=\mathbf{H}(t)\mathbf{c}%
(t),  \label{Schrodinger-2SS}
\end{equation}%
where the Hamiltonian in the rotating-wave approximation (RWA) reads \cite%
{B.Shore}
\begin{equation}
\mathbf{H}(t)=\tfrac{1}{2}\hbar \left[
\begin{array}{cc}
-\Delta (t) & \Omega (t) \\
\Omega (t) & \Delta (t)%
\end{array}%
\right] .  \label{H2}
\end{equation}%
The detuning $\Delta =\omega _{0}-\omega $ is the difference between the
transition frequency $\omega _{0}$ and the carrier laser frequency $\omega $%
. The time-varying Rabi frequency $\Omega (t)=\left\vert dE(t)\right\vert
/\hbar $ describes the laser-atom interaction, where $d$ is the electric
dipole moment for the $\psi _{1}\leftrightarrow \psi _{2}$ transition and $%
E(t)$ is the laser electric field envelope.


\subsection{Adiabatic basis}


For the derivation of the transition probability we shall need the adiabatic
basis, i.e. the basis of the eigenstates of the Hamiltonian (\ref{Ha}). We
summarize below the basic definitions and properties of this basis.

The probability amplitudes in the diabatic and adiabatic bases are connected
via the rotation matrix
\begin{equation}
\mathsf{R}(\vartheta )=\left[
\begin{array}{cc}
\cos \vartheta & \sin \vartheta \\
-\sin \vartheta & \cos \vartheta%
\end{array}%
\right] ,
\end{equation}%
as
\begin{equation}
\mathbf{c}(t)=\mathsf{R}(\vartheta (t))\mathbf{a}(t),  \label{c=Ra}
\end{equation}%
where the column-vector $\mathbf{a}(t)=[a_{-}(t),a_{+}(t)]^{T}$ comprises
the probability amplitudes of the adiabatic states $|\varphi _{-}(t)\rangle $
and $|\varphi _{+}(t)\rangle $. These amplitudes satisfy the transformed Schr%
\"{o}dinger equation,
\begin{equation}
i\hbar \frac{d}{dt}\mathbf{a}(t)=\mathsf{H}_{a}(t)\mathbf{a}(t),
\end{equation}%
where the transformed Hamiltonian is given by
\begin{eqnarray}
\mathsf{H}_{a}(t) &=&\mathsf{R}^{-1}(\vartheta (t))\mathsf{H}(t)\mathsf{R}%
(\vartheta (t))-i\hbar \mathsf{R}^{-1}(\vartheta (t))\mathsf{\dot{R}}%
(\vartheta (t))  \label{Ha} \\
&=&\hbar \left[
\begin{array}{cc}
\mathcal{E}_{-}(t) & -i\dot{\vartheta}(t) \\
i\dot{\vartheta}(t) & \mathcal{E}_{+}(t)%
\end{array}%
\right] ,
\end{eqnarray}%
where the overdots denote time derivatives. For the reader convenience we
write the expresion for the nonadiabatic couplig $\dot{\vartheta}(t)$ in
terms of $\Omega (t)$ and $\Delta (t)$. Using the definition given by Eq.(%
\ref{theta}) easely can be seen that

\begin{equation}
\dot{\vartheta}(t)=\frac{\dot{\Omega}(t)\Delta (t)-\dot{\Delta}(t)\Omega (t)%
}{\Omega ^{2}(t)+\Delta ^{2}(t)}  \label{non-ad-coup}
\end{equation}

In terms of the mixing angle $\vartheta (t)$, defined as

\begin{equation}
\tan 2\vartheta (t)=\frac{\Omega (t)}{\Delta (t)},\text{\qquad }(0\leqq
\vartheta (t)\leqq \frac{\pi }{2}),  \label{theta}
\end{equation}%
the eigenstates of $\mathsf{H}(t)$ read
\begin{subequations}
\label{adiabatic states}
\begin{eqnarray}
|\varphi _{-}(t)\rangle &=&\cos \vartheta (t)|\psi _{1}\rangle -\sin
\vartheta (t)|\psi _{2}\rangle ,  \label{phi-} \\
|\varphi _{+}(t)\rangle &=&\sin \vartheta (t)|\psi _{1}\rangle +\cos
\vartheta (t)|\psi _{2}\rangle .  \label{phi+}
\end{eqnarray}%
The time dependences of the adiabatic states $|\varphi _{-}(t)\rangle $ and $%
|\varphi _{+}(t)\rangle $ derive from the mixing angle $\vartheta (t)$,
whereas the bare (diabatic) states $|\psi _{1}\rangle $ and $|\psi
_{2}\rangle $ are stationary. The energies of the adiabatic states are the
eigenvalues of $\mathsf{H}(t)$,
\end{subequations}
\begin{equation}
\hbar \mathcal{E}_{\pm }(t)=\frac{\hbar }{2}\left[ \Delta \pm \sqrt{\Omega
^{2}(t)+\Delta ^{2}}\right] .
\end{equation}%
The splitting between them is given by
\begin{equation}
\hbar \mathcal{E}(t)=\hbar \mathcal{E}_{+}(t)-\hbar \mathcal{E}_{-}(t)=\hbar
\sqrt{\Omega ^{2}(t)+\Delta ^{2}}.  \label{splitting}
\end{equation}%
Hereafter we will consider level crossing models. Because the Rabi frequency
$\Omega (t)$ vanishes at large times, $\Omega (\pm \infty )=0$, and because
the detuning $\Delta (t)$ sweeps from minus to plus infinity, $\hbar \Delta
(\pm \infty )=\pm \infty $, the mixing angle $\vartheta (t)$ rotates
clockwise from $\vartheta (-\infty )=\pi /2$ to $\vartheta (+\infty )=0$,
and the composition of the adiabatic states changes accordingly.
Asymptotically, each adiabatic state becomes uniquely identified with a
single diabatic state,
\begin{subequations}
\label{diabatic-adiabatic}
\begin{eqnarray}
&&-|\psi _{2}\rangle \overset{-\infty \leftarrow t}{\longleftarrow }|\varphi
_{-}(t)\rangle \overset{t\rightarrow +\infty }{\longrightarrow }|\psi
_{1}\rangle , \\
&&\text{ \ \ \ }|\psi _{1}\rangle \overset{-\infty \leftarrow t}{%
\longleftarrow }|\varphi _{+}(t)\rangle \overset{t\rightarrow +\infty }{%
\longrightarrow }|\psi _{2}\rangle .
\end{eqnarray}%
Because of the level crossing each adiabatic state connects \emph{different}
bare states at $-\infty $ and $+\infty $. In the adiabatic limit, the system
starts in state $|\psi _{1}\rangle $ and follows the adiabatic state $%
|\varphi _{+}(t)\rangle $ to end up in state $|\psi _{2}\rangle $. Hence
adiabatic evolution and level crossing lead to complete population transfer.

It is important to note that the probability of transition in the adiabatic
basis $\mathcal{P}$ is equal to the probability of no transition in the
diabatic basis,
\end{subequations}
\begin{equation}
P=1-\mathcal{P}.  \label{PP}
\end{equation}%
We will continue with the description of the Dykhne-Davis-Pechukas (DDP)
method, which gives the adiabatic probability $\mathcal{P}$, and we shall
use Eq. (\ref{PP}) to find the diabatic probability $P$.


\subsection{Dykhne-Davis-Pechukas (DDP) approximation}


\subsubsection{A single transition point}


A useful and very accurate technique for obtaining the final transition
probabilities is the Dykhne-Davis-Pechukas method, or DDP method, first
introduced by Dykhne and given a rigorous mathematical formulation later by
Davis and Pechukas \cite{Davis76}. The basic idea of the DDP method is that,
in the adiabatic limit, the two-state coupling is universal, independent of
a given model, and the contributions to the t transition probability $P$
between the adiabatic states are given by the complex plane zeros of the
adiabatic eigenenergies, in the form of an exponential.
Dykhne-Davis-Pechukas (DDP) approximation, \cite{Davis76}, which provides
the asymptotically exact transition probability $P$ can be also used to
estimate the non-adiabatic effects \cite{Guerin}. The DDP formula reads
\begin{equation}
P\approx e^{-2{\text{Im}}D(t_{0})},  \label{DP-1}
\end{equation}%
where
\begin{equation}
D(t_{0})=\int_{0}^{t_{0}}E(t)dt  \label{D(Tc)}
\end{equation}%
is an integral over the eigenenergy splitting $E(t)$. The point $t_{0}$ is
called the transition point and it is defined as the (complex) zero of the
quasienergy splitting,
\begin{equation}
E(t_{0})=0,  \label{Tc-def}
\end{equation}%
which lies in the upper half of the complex $t$-plane (i.e., with Im$\mathrm{%
\,}t_{0}>0$). Equation (\ref{DP-1}) gives the correct asymptotic probability
for nonadiabatic transitions provided: (i) the quasienergy splitting $E(t)$
does not vanish for real $t$, including at $\pm \infty $; (ii) $E(t)$ is
analytic and single-valued at least throughout a region of the complex $t$%
-plane that includes the region from the real axis to the transition point $%
t_{0}$; (iii) the transition point $t_{0}$ is well separated from the other
quasienergy zero points (if any) and from possible singularities; (iv) there
exists a level (or Stokes) line defined by
\begin{equation}
\ {\text{Im}}D(t)={\text{Im}}D(t_{0}),  \label{stokes line}
\end{equation}%
which extends from $-\infty $ to $+\infty $ and passes through $t_{0}$.

As has been pointed out already by Davis and Pechukas \cite{Davis76}, for
the Landau-Zener model, which possesses a single transition point, the DDP
formula (\ref{DP-1}) gives the exact transition probability, not only in the
adiabatic limit but also in the general case. This amazing feature indicates
the relevance of the DDP approximation.


\subsubsection{Multiple transition points}


In the case of more than one zero points in the upper $t$-plane, Davis and
Pechukas \cite{Davis76} have suggested , that Eq.~(\ref{DP-1}) can be
generalized to include the contributions from all these $N$ zero points $%
t_{k}$ in a coherent sum. This suggestion has been later verified \cite%
{Joye93,Suominen92oc,Suominen92pra}. The generalized DDP formula has the
form
\begin{equation}
P\approx \left\vert \sum\nolimits_{k=1}^{N}\Gamma _{k}e^{\text{i}%
D(t_{k})}\right\vert ^{2},  \label{DP-N}
\end{equation}%
where the $\Gamma _{k}$ factors are defined by
\begin{equation}
\Gamma _{k}=4i\lim\limits_{t\rightarrow t_{k}}(t-t_{k})\dot{\vartheta}(t).
\label{Gamma-k}
\end{equation}%
and they usually take values $+1$ or $-1$. Here $\dot{\vartheta}(t)$
accounts for the nonadiabatic coupling between the adiabatic states, with $%
\vartheta (t)=\frac{1}{2}\tan ^{-1}\Omega (t)/\Delta (t)$.

In principle, Eq.~(\ref{DP-N}) should be used when there are more than one
zero points lying on the lowest Stokes line (the closest one to the real
axis) and should include only the contributions from these zeroes. The
contributions from the farther zeroes are exponentially small compared to
the dominant ones and may therefore be neglected.

\subsection{Adiabatic optimization for two-state system based on the DDP
method}

It is shown in \cite{Guerin} that Dykhne-Davis-Pechukas (DDP) method \cite%
{Davis76} can be used to examine\ the adiabatic limit of population transfer
in two-level models driven by a chirped laser field. In \cite{Guerin} the
final population transfer for different trajectories in the parameter space
in the adiabatic limit is analyzed.

After using the scaled time $t=\tau /\alpha $, where the parameter $1/\alpha
$ under the limit $\alpha \rightarrow \infty $ can be viewed as adiabatic
limit, we can write the new scaled Schrodinger equation Eq.(\ref{Sch-eq})
for the two-state system. In Eq.(\ref{H2}), we can parameterize the
trajectories defined from $\Omega (t)$ and $\Delta (t)$ as a function of
time, by assuming a given smooth pulse shape function $0<\Lambda (t)<1$,
which has its maximum for $t=0$. This pulse shape function is related to the
coupling by

\begin{equation}
\Omega (t)=\Omega _{0}\Lambda (t)  \label{Rabi-ts}
\end{equation}

where $\Omega _{0}$ is the two-state peak Rabi frequency. In reason to have
the constant eigenenergy splitting $E(t)$, accordingly to Eq.(\ref{Rabi-ts})
the detunig is defined by

\begin{equation}
\Delta (t)=\Delta _{0}\text{sign}(t)\sqrt{1-\Lambda ^{2}(t)}
\label{delta-ts}
\end{equation}

The parametrization given by Eq.(\ref{Rabi-ts}) and Eq.(\ref{delta-ts})
implies\

\begin{equation}
E(t)=\sqrt{\left( \Delta _{0}\right) ^{2}+\left[ \left( \Omega _{0}\right)
^{2}-\left( \Delta _{0}\right) ^{2}\right] \Lambda ^{2}(t)}  \label{E-ts}
\end{equation}

Using the DDP method, Eq.(\ref{Tc-def}) and Eq.(\ref{E-ts}) give

\begin{eqnarray}
\Lambda (t_{0}) &=&\pm i\frac{\Delta _{0}}{\sqrt{\left( \Omega _{0}\right)
^{2}-\left( \Delta _{0}\right) ^{2}}},\text{ for }\Omega _{0}>\Delta _{0}
\label{transition points-ts} \\
\Lambda (t_{0}) &=&\pm \frac{\Delta _{0}}{\sqrt{\left( \Delta _{0}\right)
^{2}-\left( \Omega _{0}\right) ^{2}}},\text{ for }\Omega _{0}<\Delta _{0}
\notag
\end{eqnarray}

For particular class of analytic functions, defined with $\Lambda (t)$, the
following condition is fulfilled

\begin{equation*}
\lim_{\text{Im}t_{0}\rightarrow \infty }\Lambda (t_{0})\rightarrow \infty
\end{equation*}

Using Eq.(\ref{transition points-ts}) for this class of models defined with $%
\Lambda (t),$ the difference between $\Omega _{0}$ and $\Delta _{0}$ to
tends to zero\ \

\begin{equation*}
\left\vert \Omega _{0}-\Delta _{0}\right\vert \rightarrow 0,
\end{equation*}

is necessary and sufficient condition the imaginary part of the transition
points to tends to infinity

\begin{equation*}
\text{Im}t_{0}\rightarrow \infty .
\end{equation*}

From Eq.(\ref{D(Tc)}) can be seen that

\begin{equation}
\underset{\left\vert \Omega _{0}-\Delta _{0}\right\vert \rightarrow 0}{\lim }%
\text{Im}D(t_{0})=\text{Im}\int_{0}^{\infty }\Delta _{0}dt\rightarrow \infty
\label{ImD(tc)-ts}
\end{equation}

Therefore, from Eq.(\ref{DP-1}) follows that the dominant nonadiabatic
correction given by the DDP formula vanishes for the level lines defined by
Eq.(\ref{Rabi-ts}), Eq.(\ref{delta-ts}) and $\Omega _{0}=\Delta _{0}.$ As
have been pointed in \cite{Guerin} in the adiabatic regime, the optimum
level lines can be seen as a boundary between decreasing and oscillating
regimes for the nonadiabatic correction.


\section{Second DDP estimation for the optimized adiabatic passage}


As we have expalined, Gu\'{e}rin \emph{et al.} \cite{Guerin} have used the
DDP method to optimize the adiabatic passage between two states, assuming
that the probability for nonadiabatic losses could be determined by the
brhaviour of the transition points $t_{k}$.

They have proposed to suppress the nonadiabatic losses altogether by
choosing the Rabi frequency $\Omega (t)$ and the detuning $\Delta (t)$ such
that there are \emph{no transition points}. This condition is obviously
fulfilled if the quasienergy splitting is \emph{constant},
\begin{equation}
\varepsilon (t)=\sqrt{\Omega (t)^{2}+\Delta (t)^{2}}=\text{const.}
\label{trajectory}
\end{equation}%
The later condition also manifests the choice of a detuning and Rabi
frequency defined with Eqs. (\ref{delta-ts}) and (\ref{Rabi-ts}). Easely can
be seen that the same detuning and Rabi frequency functions could be
parameterized as
\begin{subequations}
\begin{gather}
\Delta (t)=\Omega _{0}\sin \left[ \frac{\pi }{2}f(t)\right] ,\qquad \Omega
(t)=\Omega _{0}\cos \left[ \frac{\pi }{2}f(t)\right] ,  \label{omega-delta}
\\
-1=f(-\infty )\leqq f(t)\leqq f(\infty )=1,  \label{f(t)}
\end{gather}%
with $f(t)$ being an arbitrary monotonically increasing function with the
above property. According to the DDP method such models do not have
transition points and lead to vanishing nonadiabatic corrections. The
optimization based on DDP is not exact in the sense that it is performed by
using approximate technique. It is interesting to calculate the corrections
to this DDP optimization. Although we are able to design models that yield
according to the DDP, vanishing nonadiabatic corrections there is no way to
calculate the transition probability for these models again using DDP method
in the diabatic basis. DDP approximation comprise nonadiabatic corrections
not only from the first-order perturbation theory in the adiabatic basis but
adding the contributions from all orders via correct prefactor. Even so,
using DDP not in the diabatic but in the first adiabatic basis is
instructive. This is also a way to calculate the deviation form the DDP
optimization and to reveal the nature of the oscillation behavior for the
transition probability for models that are designed to minimize the
nonadiabatic corrections \ \cite{Guerin}. We note that the the probability
of transition in the adiabatic basis $\mathcal{P}$ is equal to the
probability of no transition in the diabatic basis and the both are related
via (\ref{PP}). Using the parametrization (\ref{omega-delta}) for the Schr%
\"{o}dinger equation in the adiabatic basis, the transformed Hamiltonian up
to phase transformation is given by

\end{subequations}
\begin{equation}
\mathsf{H}_{a}(t)=\hbar \left[
\begin{array}{cc}
\Omega _{0} & \dot{f}(t) \\
\dot{f}(t) & -\Omega _{0}%
\end{array}%
\right] ,  \label{Ha2}
\end{equation}

where the nonadiabatic coupling is given by (\ref{non-ad-coup}). Nonetheless
a particular level crossing model does not have a transition points in the
diabatic basis, in the first adiabatic basis, such a model depending on the
function $f(t)$ would have transition points. This means that applying a DDP
method in the first adiabatic basis instead of the diabatic one, one could
analyze the nonadiabatic corrections of the optimized adiabatic passage.

\subsection{Gaussian model}

As a particular example we will consider a level crossing model,

\begin{equation}
\Delta (t)=\Omega _{0}\sin \left[ \frac{\pi }{2}\text{erf}(t/T)\right]
,\qquad \Omega (t)=\Omega _{0}\cos \left[ \frac{\pi }{2}\text{erf}(t/T)%
\right] .  \label{erf}
\end{equation}%
This model in the adiabatic basis is related to the Gaussian model, which is
seen from the Hamiltonian (\ref{Ha2}).
\begin{equation*}
\text{i}\hbar \frac{\text{d}}{\text{d}t}\mathbf{a}(t)=\hbar \left[
\begin{array}{cc}
\Omega _{0} & \frac{\sqrt{\pi }}{T\text{ }}\exp \left[ -\left( t/T\right)
^{2}\right] \\
\frac{\sqrt{\pi }}{T\text{ }}\exp \left[ -\left( t/T\right) ^{2}\right] &
-\Omega _{0}%
\end{array}%
\right] \mathbf{a}(t),
\end{equation*}

The analytic estimation of the transition probability for the Gaussian model
could be performed using DDP. We will briefly review details of such
calculation, but the reader could find a similar calculation in more details
in \cite{Vasilev1}.


\subsubsection{Transition points}


For the Gaussian model (\ref{model}), there are infinitely many transition
points in the upper half-plane. 
In terms of the dimensionless time $\tau =t/T=\xi +i\eta $, they are given
by
\begin{subequations}
\label{transition points}
\begin{eqnarray}
\tau _{k}^{\pm } &=&\pm \xi _{k}+i\eta _{k},  \label{t_c} \\
\xi _{k} &=&\frac{1}{2}\sqrt{\sqrt{4\left( \ln \alpha \right) ^{2}+\left(
2k+1\right) ^{2}\pi ^{2}}+2\ln \alpha },  \label{x_c} \\
\eta _{k} &=&\frac{1}{2}\sqrt{\sqrt{4\left( \ln \alpha \right) ^{2}+\left(
2k+1\right) ^{2}\pi ^{2}}-2\ln \alpha },  \label{y_c}
\end{eqnarray}%
where $k=0,1,2,...$ and
\end{subequations}
\begin{equation}
\alpha =\frac{\sqrt{\pi }}{T\text{ }\Omega _{0}}.  \label{alpha}
\end{equation}

For $\alpha \ll 1$, we have
\begin{subequations}
\label{transition points small}
\begin{eqnarray}
\xi _{k} &\sim &\frac{(2k+1)\pi }{4\sqrt{\ln (1/\alpha )}},\qquad (\alpha
\ll 1),  \label{x_c small} \\
\eta _{k} &\sim &\sqrt{\ln (1/\alpha )},\qquad (\alpha \ll 1).
\label{y_c small}
\end{eqnarray}%
Hence, as $\alpha $ decreases, the transition points approach the imaginary
axis and in the limit $\alpha \rightarrow 0$ coalesce (logarithmically) with
their counterparts in the second quadrant.

As we have mentioned the transition probability for the models choosen to
satisfy the DDP optimization condition (\ref{trajectory}) show oscillating
behavior, although they should\ yield optimized addiabatic passage. This is
due to the fact that DDP is approximate method. A particular model without
transition points in the diabatic basis, generally has a transition point in
the first adiabatic basis. As in the Gaussian model the contributions from
this transition points lead to oscilations, according to DDP formula (\ref%
{DP-N}). It is important to note the relation between the asymptotic
bahaviour of the transition points for $\alpha \ll 1$ and the adiabatic
limit. From the definition of the model (\ref{erf}) is clear that the limit $%
T\rightarrow \infty $ can be seen as the adiabatic limit, so $1/T$ play role
of the adiabatic parameter. From the definition (\ref{alpha}) we see that $%
\alpha $ is proportional to the adiabatic parameter and the limit $\alpha
\rightarrow 0$ is the adiabatic limit. According to the asymptotic behavior
of the transition points, the transition point from the first quadrant
logarithmically coalesce with their counterpart in the second quadrant.and
approach the imaginary axis. Since in the limit $\alpha \rightarrow 0$,
which is perfect adiabatic regime, we do not have a coherent contribution to
the DDP formula (\ref{DP-N}) from two transition points, no oscillations
will be seen. This simple analysis shows that even in the first adiabatic
basis, the correct asymptotic behaviors of the optimized adiabatic passage
is achieved.


\subsubsection{DDP integrals}


Because for the Gaussian model (\ref{model}) there are infinitely many
transition points, the most accurate transition probability is expected to
be given by the generalized DDP formula (\ref{DP-N}). The dominant
contributions to the sum in this formula originate from the two transition
points closest to the real axis, $\tau_0^-$ and $\tau_0^+$. For simplicity,
we neglect the contributions from all others and retain only the terms from
these two points.

Because $\left( \tau_{0}^{-}\right) ^{\ast }=-\tau_{0}^{+}$ and because $%
\mathcal{E}(\tau)$ is an even function of time, it is easy to show that
\end{subequations}
\begin{equation}
\mathcal{D}(\tau_{0}^{-})=-\mathcal{D}^{\ast }(\tau_{0}^{+}),  \label{D-D}
\end{equation}
that is ${\text{Re}}\mathcal{D}(\tau_{0}^{-})=-{\text{Re}}\mathcal{D}%
(\tau_{0}^{+})$ and ${\text{Im}}\mathcal{D}(\tau_{0}^{-})={\text{Im}}%
\mathcal{D}(\tau_{0}^{+})$. Hence it is sufficient to calculate only one of
these integrals and we choose $\mathcal{D}(\tau_{0}^{+})$ for this purpose.

Because the imaginary part of the DDP integral $\mathcal{D}(\tau)$ is the
same for the two transition points $\tau_0^+$ and $\tau_0^-$ [cf.~Eq.~(\ref%
{D-D})], these points lie on the same Stokes line, defined by Eq.~(\ref%
{Stokes line}). This Stokes line extends from $-\infty$ to $+\infty$, which
is a necessary condition for the validity of the DDP approximation \cite%
{Davis76,Joye91}.

With the arguments presented above, the problem is reduced to the
calculation of the DDP integral
\begin{equation}
\mathcal{D}(\tau_{0}^{+}) = \Delta T\int_{0}^{\tau_{0}^{+}}\sqrt{%
\alpha^{2}e^{-2\tau^2}+1}\,d\tau .  \label{D(Tc)}
\end{equation}
The estimation of this integral will be our main concern hereafter in this
section.


\paragraph{Asymptotic behavior of the DDP integral for small $\protect\alpha
$}


For small $\alpha $ ($\alpha \ll 1$) we expand the integrand in Eq. (\ref%
{D(Tc)}) by using the Taylor expansion, and perform term-by-term
integration. This integration is justified within the circle $\left\vert
x\right\vert \leqq 1$, where the series (\ref{sqrt(1+x)}) is uniformly
convergent. We choose the path of integration to be the straight line from $%
\tau =0$ to $\tau =\tau _{0}^{+}$ and parameterize this path as $\tau =\tau
_{0}^{+}s$ ($0\leqq s\leqq 1$). It is easy to see that $|\alpha
^{2}e^{-2\tau ^{2}}|\leqq 1$ along this path. Indeed,
\begin{equation*}
|\alpha ^{2}e^{-2\tau ^{2}}|=\alpha ^{2}|e^{-2(\tau
_{0}^{+})^{2}s^{2}}|=\alpha ^{2(1-s^{2})}\leqq 1,
\end{equation*}%
because $\alpha <1$ and $0\leqq s\leqq 1$.

By using the relation
\begin{equation}
\int_{0}^{\tau _{0}^{+}}e^{-2nu^{2}}du=\frac{\sqrt{\pi }\text{Erf}(\tau
_{0}^{+}\sqrt{2n})}{2\sqrt{2n}},
\end{equation}%
we find that
\begin{eqnarray}
\mathcal{D}(\tau _{0}^{+}) &=&\Delta T\left[ \tau
_{0}^{+}+\sum_{n=1}^{\infty }(-1)^{n-1}\frac{(2n-3)!!}{(2n)!!}\right.  \notag
\\
&&\times \left. \frac{\alpha ^{2n}\sqrt{\pi }\text{Erf}(\tau _{0}^{+}\sqrt{2n%
})}{2\sqrt{2n}}\right] ,  \label{D-small-series}
\end{eqnarray}


\subsubsection{Uniform approximation to the DDP integral}


It is shown in \cite{Vasilev1}, one can derive derive a uniform
approximation to the DDP integral $\mathcal{D}(\tau _{0}^{+})$, Eq.~(\ref%
{D(Tc)}), by choosing an appropriate integration contour.

\paragraph{Imaginary part of the DDP integral}


For the imaginary part of the DDP integral (\ref{D(Tc)}) we have
\begin{equation}
{\text{Im}}\mathcal{D}(\tau _{0}^{+})\approx \frac{1}{2}\Delta T\sqrt{\sqrt{%
4\ln ^{2}(m\alpha )+\pi ^{2}}-2\ln (m\alpha )}.  \label{ImD}
\end{equation}
and $m\approx 1.311468$.

The advantage of this choice is that the approximation (\ref{ImD}), besides
providing the exact result for $\alpha =1$ is also very accurate in some
vicinity of this important point. On the other hand, Eq. (\ref{ImD}) has the
following asymptotics
\begin{subequations}
\label{ImD-limit}
\begin{equation}
{\text{Im}}\mathcal{D}(\tau _{0}^{+})\sim \Delta T\sqrt{\ln (m/\alpha )}%
,\qquad (\alpha \ll 1),
\end{equation}%
These expressions agree with Eqs. (\ref{ImD-small}) and (\ref{ImD-large}),
except for the factor $m$, which is insignificant in the limits $\alpha \gg
1 $ and $\alpha \ll 1$ [since $\ln (m\alpha )=\ln m+\ln \alpha \approx \ln
\alpha $ for $\alpha \gg 1$ and similarly for $\alpha \ll 1$]. This factor
becomes significant for intermediate $\alpha $, where, however, the accuracy
of Eq. (\ref{ImD}) improves until, as explained above, it becomes exact for $%
\alpha =1$.

\paragraph{Real part of the DDP integral}


For the real part of the DDP integral (\ref{D(Tc)}) we have
\end{subequations}
\begin{equation}
{\text{Re}}\mathcal{D}(\tau _{0}^{+})=\Delta T\left[ \mathcal{I}_{1}(\alpha
)+\mathcal{I}_{2}(\alpha )\right] .
\end{equation}

The integral $\mathcal{I}_{1}(\alpha )$ is approximated as
\begin{equation}
\mathcal{I}_{1}(\alpha )\approx \left( \sqrt{\alpha ^{2}+1}-1\right) \sqrt{%
\frac{1}{2}\ln \frac{\alpha ^{2}}{\left[ 1+\nu \left( \sqrt{\alpha ^{2}+1}%
-1\right) \right] ^{2}-1}}.
\end{equation}%
and $\nu =0.462350...$The second integral $\mathcal{I}_{2}(\alpha )$.is
approximated as
\begin{equation}
\mathcal{I}_{2}(\alpha )\approx =\frac{1}{2}\sqrt{\sqrt{\left[ \ln \frac{%
\alpha ^{2}}{\mu (2-\mu )}\right] ^{2}+\pi ^{2}}+\ln \frac{\alpha ^{2}}{\mu
(2-\mu )}},
\end{equation}%
where $\mu =0.316193...$, and $\mu (2-\mu )=0.532408...$

\subsubsection{Transition probability}


In order to sum the contributions from various DDP integrals we need the
factors $\Gamma _{k}$, Eq. (\ref{Gamma-k}). One finds after simple algebra
that
\begin{equation}  \label{Gamma-k-1}
\Gamma(\tau_k^{\pm}) = \pm(-1)^k.
\end{equation}

Collecting the results we find
\begin{equation}
\mathcal{P}\sim 4\exp \left[ -2{\text{Im}}\mathcal{D}(\tau _{0}^{+})\right]
\sin ^{2}\left[ {\text{Re}}\mathcal{D}(\tau _{0}^{+})\right] .  \label{P-DDP}
\end{equation}%
In \cite{Vasilev1}, is shown that including the contributions from all
transition points one can verify the following expresion for the transition
probability
\begin{equation}
\mathcal{P}\sim \frac{\sin ^{2}\left[ {\text{Re}}\mathcal{D}(\tau _{0}^{+})%
\right] }{\cosh ^{2}\left[ {\text{Im}}\mathcal{D}(\tau _{0}^{+})\right] }.
\label{P-DDP-sech}
\end{equation}

\begin{figure}[tb]
\includegraphics[width=75mm]{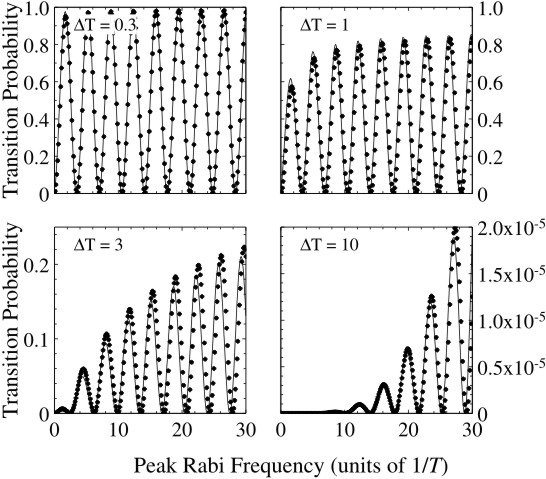}
\caption{Transition probability for the Gaussian pulse plotted vs
the peak
 Rabi frequency $\Omega_0$ for four values of the detuning, $\Delta T=0.3,1,3,10$.
The exact results obtained by numerical integration of the
Schr\"{o}dinger equation are shown by dots
 and the approximation by solid lines.}
\label{Fig-Rabi}
\end{figure}

Equation (\ref{P-DDP-sech}) provides a very accurate description
of the transition probability $\mathcal{P}$. This approximation is
plotted on Fig. \ref{Fig-Rabi} as a function of the peak Rabi
frequency $\Omega_{0}$ for four different values of the detuning
$\Delta$. As $\Omega_{0}$ increases, Rabi-like oscillations are
observed, with both amplitude and frequency matched very well by
our approximation (\ref{P-DDP-sech}).


\subsection{Deviation form optimizaed pulses}

The DDP based approximation for the transition probability can be derived in
the case of absent transition points by using the same DDP technique but in
the first adiabatic basis. A very reasonable question is how we can derived
transition probability whenever a small deviation from optimized pulses
takes place.Insted of the optimized pulses given by Eq.(\ref{erf}) we
introduce

\begin{equation}
\Delta (t)=\left( \Omega _{0}+\mu \right) \sin \left[ \frac{\pi }{2}\text{erf%
}(t/T)\right] ,\qquad \Omega (t)=\Omega _{0}\cos \left[ \frac{\pi }{2}\text{%
erf}(t/T)\right]  \label{erf-mu}
\end{equation}

where $\mu $ stands for small parameter. If we apply DDP for such model, for
$\mu =0$ as have been shown DDP fails to describe transition probability.
This is true not only for $\mu =0$ but for some vicinity of this point.

\begin{figure}[tb]
\includegraphics[width=75mm]{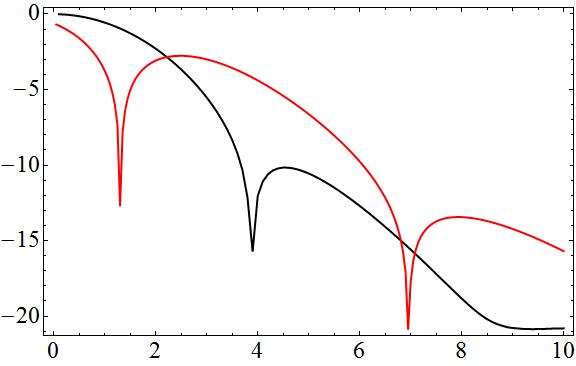}
\caption{$\ln \left( 1-\mathcal{P}\right) $ as a function of Rabi
frequency $\Omega _{0}$. Black curve shows the
pulses given in Eq.(\ref{erf}), Red curve shows pulses given with Eq.(\ref%
{erf-mu}) for $\mu =1$} \label{probgr}
\end{figure}

Figure \ref{probgr} displays in Log scale 1 -  transition probability $%
\mathcal{P}$ i.e. $\ln \left( 1-\mathcal{P}\right) $ for the adiabatic
optimizad pulses given in Eq.(\ref{erf}) and pulses given with Eq.(\ref%
{erf-mu}) as a function of peak Rabi frequency $\Omega _{0}$.

\section{Conclusions\label{Sec-conclusion}}


We have examined the optimization via DDP method in the superadibatic basis.
According to the DDP method models that do not have transition points would
lead to vanishing nonadiabatic corrections. This would be the essence of the
optimization based on DDP technique. We have shown that this condition is
not sufficient for perfect adiabtic optimization. This is due to the
approximate origin of the DDP method itself. Nevertheless DDP has been
derived in order to take into account higher order adiabatic corrections,
even for models that should yield according to DDP perfect adiabatic
evolution(i.e. there are no transition points) within the the next adiabatic
basis consecutive optimization is possible.

\acknowledgments This work has been supported by the project QUANTNET -
European Reintegration Grant (ERG) - PERG07-GA-2010-268432.


\end{document}